\documentclass[conference]{IEEEtran}
\usepackage{dsfont}
\usepackage{amssymb}
\usepackage{subfig}
\usepackage{graphicx, amssymb, amsmath}
\usepackage{extarrows}
\emergencystretch=\maxdimen
\IEEEoverridecommandlockouts
\begin{document}

\title{Spatial Traffic Shaping in Heterogeneous Cellular Networks with Energy Harvesting}

\author{\IEEEauthorblockN{Shan~Zhang, Sheng~Zhou, Jie Gong, Zhisheng~Niu}
\IEEEauthorblockA{\\ %Tsinghua National Laboratory for Information Science and Technology,\\
Dept. of Electronic Engineering, \\ Tsinghua Univ., Beijing, 100084, P.R. China \\
Email: \{zhangshan11\}@mails.tsinghua.edu.cn, \\ \{sheng.zhou, gongj13, niuzhs\}@tsinghua.edu.cn}
\and
\IEEEauthorblockN{Ning~Zhang, Xueming~(Sherman)~Shen}
\IEEEauthorblockA{\\Department of Electrical and Computer Engineering, \\ University of Waterloo, 200 University Avenue West, \\ Waterloo, Ontario, Canada, N2L 3G1 \\ Email: n35zhang@uwaterloo.ca, xshen@bbcr.uwaterloo.ca}
\thanks{This work is sponsored in part by the National Basic Research Program of China (973 Program: No. 2012CB316001), the National Science Foundation of China (NSFC) under grant No. 61201191, No. 61321061, No. 61401250, and No. 61461136004, and Hitachi Ltd..
%This work is sponsored in part by the National Basic Research Program of China (973 Program: No. 2012CB316001), the National Science Foundation of China (NSFC) under grant No. 61201191, No. 61321061, No. 61401250, and No. 61461136004, and Intel Collaborative Research Institute for Mobile Networking and Computing¡£
}}

\maketitle
\begin{abstract}
Energy harvesting (EH), which explores renewable energy as a supplementary power source, is a promising 5G technology to support the huge energy demand of heterogeneous cellular networks (HCNs).
However, the random arrival of renewable energy brings new challenges to network management.
Through adjusting the distribution of traffic load in spatial domain, traffic shaping helps to balance the cell-level power demand and supply, enhancing the utilization of renewable energy.
In this paper, we investigate the power saving performance of traffic shaping in an analytical way.
Specifically, an energy-optimal traffic shaping scheme (EOTS) is devised for HCNs with EH, whereby the on-off state of the off-grid small cell and the amount of offloading traffic are adjusted dynamically, based on the statistic information of energy arrival and traffic load.
Numerical results are given, which show that EOTS scheme can significantly reduce the on-grid power consumption for the daily traffic and solar energy profiles, compared with the greedy method where users are always offloaded to the off-grid small cell with priority.

\end{abstract}

\section{Introduction}

To accommodate the exponential increasing traffic demand, small cells (SCs) are expected to be densely deployed overlaid with the conventional macro base stations (MBSs), introducing the heterogeneous cellular network (HCN) architecture.
With high network capacity, HCNs will play an important role for 5G evolution \cite{NZhang_cloud_5G}.
However, the huge energy demand and costly deployment of SCs will cause heavy burdens to the network operators.
To effectively reduce network energy consumption, dynamic network planning has been proposed and extensively investigated, whereby base stations can be dynamically turned off with traffic variations \cite{mine_Globecom_load}-\cite{mine_TWC}.

In addition to energy saving methods, energy harvesting (EH) technology is also introduced into HCNs, whereby new types of SCs equipped with EH devices (like solar panels or wind turbines) can exploit renewable energy for power source \cite{Yuyi_Magazine_RE}.
In addition, together with mmWave wireless backhaul, EH can remove the wired connection of SCs, enabling flexible and cost-efficient network deployment.
Whereas, the random arrival of renewable energy poses significant challenges, and the design of efficient network management schemes is a key issue for HCNs with EH \cite{Yuyi_Magazine_RE}.

Plenty of works have been done to improve the utilization of harvested energy \cite{EH_energy_coop_2_BS}-\cite{BUPT_conf_Het_RE}.
Energy cooperation among BSs adjusts the renewable energy supply of each BS to match the traffic demand, which can be realized through dedicated power line connections between BSs \cite{EH_energy_coop_2_BS} or the wireless energy transfer technology \cite{EH_energy_coop_wiereless_2}.
However, it is costly to deploy dedicated wired connections among BSs, and the efficiency of wireless energy transfer is still very limited.
BS cooperation can also be leveraged for traffic shaping, which adjusts the traffic distribution in spatial domain to match the energy status \cite{Tao_ICE} \cite{BUPT_conf_Het_RE}.
For the single-tier homogeneous network jointly powered by EH and power grid, traffic shaping is realized by adjusting the cell size, with the objective to minimize the on-grid power consumption \cite{Tao_ICE}.
In addition, the traffic shaping scheme based on dynamic programming is proposed for an HCN consisting of a conventional MBS and a SC jointly powered by the power grid and EH, where users are dynamically offloaded from the MBS to the SC according to the traffic and energy status \cite{BUPT_conf_Het_RE}.

In this work, the power saving performance of traffic shaping is investigated in an analytical way.
We consider an HCN covered by a conventional on-grid MBS, where one off-grid SC is further deployed to enhance the network capacity through EH.
Traffic shaping, which offloads traffic from the MBS to the SC opportunistically based on the renewable energy status, can improve the utilization of harvested energy while guaranteeing the quality of service (QoS) (i.e., rate outage probability).
Under the semi-dynamic traffic and energy models, we derive the closed-form on-grid power saving gain of traffic shaping, where the additional cost caused by user handovers is also considered.
Based on the derived power saving gain, an energy-optimal traffic shaping (EOTS) scheme is proposed, whereby the SC is dynamically switched on/off and the amount of traffic offloaded to the SC is also adjusted, with the variation of energy arrival rate.
For the daily traffic and solar energy profiles, numerical results show that the EOTS can significantly reduce the on-grid power consumption, compared with the greedy method where users are always offloaded to the off-grid small cell with priority.
The main contribution of this work is that the proposed EOTS scheme offers insights for real network operations, e.g., whether the SCs should be activated and how much traffic should be offloaded to the SCs.

The rest of the paper is organized as follows. Section~\ref{sec_system_model} introduces the system model, followed by the analysis of the power demand and supply in Section~\ref{sec_power_demand_supply}. Then, the EOTS scheme is proposed in Section~\ref{sec_Single_SC}, based on the derived on-grid power saving gain. Numerical results are presented in Section~\ref{sec_numerical_results}. Finally, we conclude the paper in Section~\ref{sec_conclusions}.
    %%%%%%%%%%%%%%%%%%%%%%%%%%%%%%

\section{System Model}
    \label{sec_system_model}
    \subsection{HCNs with Hybrid Energy Supply}

\begin{figure}
    \centering
  \includegraphics[width=3.0in]{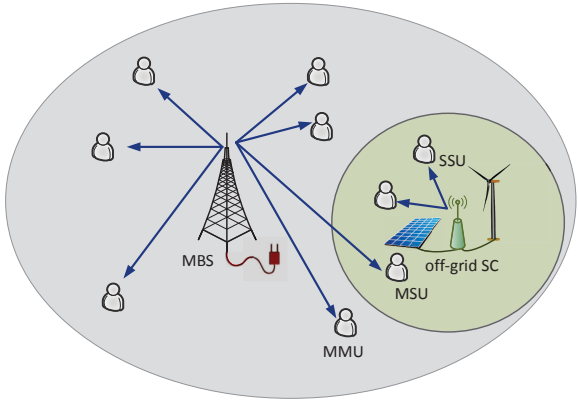}\\
  \caption{Illustration of a HCN with various energy sources.}\label{fig_HCN}
\end{figure}

        Consider a HCN whose basic coverage is guaranteed by a conventional on-grid MBS, and an off-grid SC is further deployed to enhance the network capacity with harvested energy, as shown in Fig.~\ref{fig_HCN}.
        Denote the coverage radius of MBS and SC as $D_\mathrm{m}$ and $D_\mathrm{s}$, respectively.
        We model the user distribution as a non-homogeneous Poisson Point Process (PPP) in the spatial domain, whose density is $\rho_\mathrm{m}(t)$ (outside of the SC range) and $\rho_{\mathrm{s}}(t)$ (inside of the SC) at time $t$.
        Users within the SC can be either served by the MBS or offloaded to the SC for power saving, whereas users located outside of the SC range are all associated with the MBS.
        Thus, all users can be classified into three types: (1) MMUs, located outside of the SC and served by the MBS; (2) MSUs, located in the SC but served by the MBS; and (3) SSUs, offloaded to the SC.
        According to the properties of PPP \cite{Stochastic_Geometry}, the distributions of SSUs and MSUs within the SC should also follow PPP with density $\varphi(t) \rho_{\mathrm{s}}(t)$ and $(1-\varphi(t)) \rho_{\mathrm{s}}(t)$ respectively, if users located within the SC are offloaded to the SC randomly with probability $\varphi(t)$ (i.e., random offloading scheme).
        Note that the offloading probability $\varphi(t)$ reflects the amount of traffic offloaded to the SC, which should be adjusted according to the energy arrival rate.

        Assume the MBS and SC use orthogonal bandwidth to avoid the cross-tier interference.
        The bandwidths for the MBS and SC are denoted as $W_\mathrm{m}$ and $W_\mathrm{s}$ respectively, which can be totally or partially utilized according to the traffic demand.
        Denote by $w_{\mathrm{ss}}\leq W_\mathrm{s}$ the utilized bandwidth at the SC, which is equally allocated to all SSUs.
        As for the MBS, $w_\mathrm{mm}$ is utilized to serve MMUs, while extra bandwidth $w_{\mathrm{ms}}$ is shared by MSUs.
        To guarantee the QoS, $w_\mathrm{mm}$ and $w_{\mathrm{ms}}$ are adapted to traffic and energy variations, satisfying $w_\mathrm{mm} + w_{\mathrm{ms}} \leq W_\mathrm{m}$.

        Semi-dynamic traffic and energy models are adopted, and the time line is divided into $T$ periods.
        In each period, the user density and energy harvesting rate are assumed to be static, but may change over periods.
        We solve the problem in a specific period $t$, and the subscript $t$ is omitted in the following.

    \subsection{Wireless Communication Model}

        Denote by $P_{\mathrm{Tm}}$ and $P_\mathrm{Ts}$ the transmit power of the MBS and SC, respectively.
        If user $u$ is served by the MBS, the received signal to interference plus noise ratio (SINR) is as follows:
            \begin{equation}\label{eq_SINR_MBS}
                \gamma_{\mathrm{m}} = \frac{P_{\mathrm{Tm}} w_u}{W_\mathrm{m}} \frac{ {d_{\mathrm{m},u}}^{-\alpha_\mathrm{m}} h_{\mathrm{m},u}} { (\theta_\mathrm{m}+1) (\sigma^2 w_u) },
            \end{equation}
        where $w_u$ denote the bandwidth allocated to user $u$, $d_{\mathrm{m},u}$ is the transmission distance, $\alpha_\mathrm{m}$ is the path loss exponent of MBSs, $h_{\mathrm{m},u}$ reflects Rayleigh fading (i.e., $h_{\mathrm{m},u}$ follows exponential distribution with mean 1), $\sigma^2$ denotes the density of addictive Gaussian noise,  and $\theta_\mathrm{m} (\sigma^2 w_u)$ equals to the interference among MBSs.
        For MMUs and MSUs, the achievable rates are as follows
            \begin{equation}\label{eq_r_m}
                \begin{split}
                r_{\mathrm{mm}} & = \frac{w_{\mathrm{mm}}}{K_{\mathrm{mm}}+1} \log_2 (1 + \gamma_{\mathrm{m}}), ~~\mbox{for MMUs},\\
                r_{\mathrm{ms}} & = \frac{w_{\mathrm{ms}}}{K_{\mathrm{ms}}+1} \log_2 (1 + \gamma_{\mathrm{m}}), ~~\mbox{for MSUs},
                \end{split}
            \end{equation}
        where and $K_\mathrm{mm}$ and $K_{\mathrm{ms}}$ denote the number of residual MMUs and MSUs (except user $u$) , respectively.

        If user $u$ is offloaded to the SC, the received SINR is
            \begin{equation}\label{eq_SINR_SC}
                \gamma_{\mathrm{ss}} = \frac{P_{\mathrm{Ts}} w_u}{W_\mathrm{s}} \frac{ {d_{\mathrm{s},u}}^{-\alpha_\mathrm{s}} h_{\mathrm{s},u}} { \left( \theta_\mathrm{s} +1 \right) (\sigma^2 w_u) },
            \end{equation}
        where $d_{\mathrm{s},u}$ represents the transmission distance, $\alpha_\mathrm{s}$ is the path loss exponent of SCs, $h_{\mathrm{s},u}$ reflects the effect of Rayleigh fading, and $\theta_\mathrm{s} (\sigma^2 w_u) $ equals to the interference.
        Since SSUs share $w_\mathrm{ss}$ equally, the achievable data rate of SSU $u$ is given by:
            \begin{equation}\label{eq_r_s}
                r_{\mathrm{ss}} = \frac{w_{\mathrm{ss}}}{K_{\mathrm{ss}}+1} \log_2 (1 + \gamma_{\mathrm{ss}}),
            \end{equation}
        where $K_{\mathrm{ss}}$ denotes the number of residential SSUs.

%%%%%%%%%%%%%%%%%%%%%%%%%%%%%%%%%%%%%%%%%%%%%%%%%%%%%%%%%%%%%%%%%%%%%%%%%%%%%%%%%%%%%%%%%%%%%%%%%%%
\section{Analysis of Power Supply and Consumption}
    \label{sec_power_demand_supply}

    \subsection{BS Power Consumption}

        %BSs can work in either \emph{active mode} or \emph{sleep mode}, with different power consumption parameters.
        When a BS is working in \emph{active mode}, its power consumption is given by \cite{EARTH}
            \begin{equation}\label{eq_P_BS_2}
                P_\mathrm{BS} = P_0 + \frac{w}{W} \beta P_\mathrm{T},
            \end{equation}
        where $P_0$ is the static power consumed by devices like the air conditioner, $\frac{1}{\beta}$ reflects the power amplifier efficiency, $P_\mathrm{T}$ denotes the transmit power level, $W$ is the system bandwidth $w$ is the bandwidth utilized.
        If the BS is turned off (i.e., \emph{sleep mode}), the power consumed $P_\mathrm{s}$ is negligible compared with $P_0$, and thus $P_\mathrm{s}$ is approximated as zero. %\cite{LTE_standard}.
        In this work, we consider a constant power level %according to the LTE standard \cite{LTE_standard}
        (i.e., $P_\mathrm{T}$ is a constant), while the power consumption is controlled by adjusting the utilized bandwidth $w$.
        %System parameters like $P_0$, $\beta$, and $P_\mathrm{T}$ depends on the type of BSs \cite{EARTH}.

\begin{figure}
  \centering
  \includegraphics[width=3.0in]{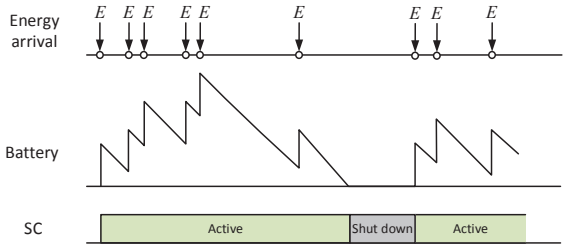}\\
  \caption{Renewable energy arrival and consumption process.}\label{fig_work_mode}
\end{figure}

    \subsection{Energy Queue Analysis}

        We apply discrete energy model to describe the EH process, with a unit of energy denoted as $E$. %\cite{xiaoxia_EH_d2d}.
        Renewable energy arrival follows Poisson process with rate $\lambda_\mathrm{E}$, which is saved in the battery for future use.
        The battery capacity is considered to be sufficiently large (i.e., no battery overflow).
        Then, the battery status of the SC is illustrated in Fig.~\ref{fig_work_mode}, with users offloaded to the SC opportunistically when the battery is not empty.
        Note that the renewable energy supply and consumption process can be modeled as an M/D/1 queue, with arrival rate $\lambda_\mathrm{E}$ and service rate $\mu_\mathrm{E}$.
        The physical meaning of $\mu_\mathrm{E}$ is the \emph{Energy Consumption Rate} of the SC, which depends on the utilized bandwidth $w_\mathrm{ss}$ according to Eq.~(\ref{eq_P_BS_2}):
            \begin{equation}\label{eq_mu}
                \mu_{\mathrm{E}} = \frac{1}{E} \left(P_{0s} + \frac{w_\mathrm{ss}}{W_\mathrm{s}}\beta_\mathrm{s} P_{\mathrm{Ts}} \right),
            \end{equation}
        where the subscript ``s'' denotes ``SC''.

        Next, we analyze the renewable energy queue.
        If $\frac{\lambda_\mathrm{E}}{\mu_\mathrm{E}} \geq 1$, the energy queue is not stable and the queue length will go to infinity.
        In this case, the harvested energy is sufficient, and the SC can be always active for traffic offloading.
        Otherwise, the SC may be shut down due to empty battery, and meanwhile, the SSUs will go back to the MBS for service.
        The corresponding probability can be obtained according to queueing theory \cite{MD1_queue}:
            \begin{equation}\label{eq_MD1_steady}
                    p_\mathrm{off}  =  1-\frac{\lambda_\mathrm{E}}{\mu_\mathrm{E}}.
            \end{equation}
        Notice that handover procedures are conducted at the moments of shutdown and reactivation, which brings additional signaling overhead and power consumption.
        The SC is shut down when the energy queue length transits from 1 to 0, with frequency $e^{-\frac{\lambda_\mathrm{E}}{\mu_\mathrm{E}}} \mu_\mathrm{E}$ \cite{MD1_queue}.
        Considering the duality between operations of shutdown and reactivation, the additional handover power consumption $P_\mathrm{ho}$ can be derived:
            \begin{equation}\label{eq_RSC_switching}
                    P_\mathrm{ho} = 2 C_\mathrm{ho} p_1 e^{-\frac{\lambda_\mathrm{E}}{\mu_\mathrm{E}}} \mu_\mathrm{E}, \\
            \end{equation}
        where $C_\mathrm{ho}$ is in J denoting the energy consumption of one handover process, $p_1$ denotes the probability that the energy queue length is 1.
        For the M/D/1 queue, $p_1$ is given by $p_1=(1-\frac{\lambda_\mathrm{E}}{\mu_\mathrm{E}}) \left(e^{\frac{\lambda_\mathrm{E}} {\mu_\mathrm{E}}} - 1\right)$ \cite{MD1_queue}.

%%%%%%%%%%%%%%%%%%%%%%%%%%%%%%%%%%%%%%%%%%%%%%%%%%%%%%%%%%%%%%%%%%%%%%%%%%%%%%%%%%%%%%%%%%%%%%%%%%%
\section{Power Saving Gain of Traffic Offloading}
    \label{sec_Single_SC}
    % In this section, we derive the on-grid power saving gain by offloading users from the MBS to the SC.
    Motivated by the power consumption model Eq.~(\ref{eq_P_BS_2}), the bandwidth required to satisfy the outage probability constraints should be obtained for power consumption analysis.
    In this section, the outage probabilities of different users are derived first, based on which we obtain the on-grid power consumption gain through traffic offloading.
    Furthermore, the EOTS scheme is proposed, which dynamically turns on/off the SC and adjusts the amount of traffic offloaded to the SC.

    \subsection{Outage Probability Analysis}

    Denote $G_\mathrm{ss}$ and $G_\mathrm{ms}$ the outage probabilities of SSUs and MSUs, respectively.
    We apply Lemmas~1 and 2 to simplify the outage probability constraints $G_\mathrm{ss}\leq \eta$ and $G_\mathrm{ms}\leq \eta$.

    \textbf{Lemma~1.} If $\frac{\sigma^2 W_\mathrm{s}}{P_{\mathrm{Ts}}}\rightarrow 0$ (high signal to noise ratio (SNR)) and $\frac{R_\mathrm{th}}{w_{\mathrm{ss}}} \rightarrow 0$ (sufficient bandwidth), the service outage constraint of SSUs $G_\mathrm{ss}=\mathds{P}\{r_\mathrm{ss}<R_\mathrm{th}\}\leq \eta$ is equivalent to
        \begin{equation}\label{eq_outage_SC_simple}
            \bar{w}_{\mathrm{ss}} \tau_{\mathrm{ss}} \geq R_\mathrm{th},
        \end{equation}
    where $R_\mathrm{th}$ is the required data rate, $\bar{w}_{\mathrm{ss}}=\frac{w_{\mathrm{ss}}}{1+ \varphi \rho_{\mathrm{s}} \pi {D_{\mathrm{s}}}^2}$ is the expected bandwidth allocated to each SSU, and $\tau_{\mathrm{ss}}$ denotes the spectrum efficiency of cell edge users:
        \begin{equation}\label{eq_tau_SC}
            \tau_{\mathrm{ss}} = \log_2 \left( 1+ \frac{P_{\mathrm{Ts}}}{\sigma^2 W_\mathrm{s} (\theta_\mathrm{s}+1)} \frac{\alpha_\mathrm{s}+2}{2 W_\mathrm{s}} \frac{\eta}{{D_{\mathrm{s}}}^{\alpha_\mathrm{s}}} \right).
        \end{equation}
    \emph{Proof}: Please refer to Appendix~\ref{appendix_SC}.
    \hfill \rule{4pt}{8pt}

    \textbf{Lemma~2.} If $\frac{\sigma^2 W_\mathrm{m}}{P_\mathrm{Tm}}\rightarrow 0$ (high SNR) and $\frac{R_\mathrm{th}}{w_{\mathrm{ms}}}\rightarrow 0$ (sufficient bandwidth), the closed-form outage probability of the MSUs can be obtained by approximating all MSUs located at the SC. The outage probability constraint $G_{\mathrm{ms}}=\mathds{P}\{r_\mathrm{ms}<R_\mathrm{th}\} \leq \eta$ is equivalent to
        \begin{equation}\label{eq_outage_ms_simple}
                \bar{w}_{\mathrm{ms}} \tau_{\mathrm{ms}} \geq R_\mathrm{th},
            \end{equation}
    where $\bar{w}_{\mathrm{ms}}=\frac{w_{\mathrm{ms}}}{1+ (1-\varphi) \rho_{\mathrm{s}} \pi {D_{\mathrm{s}}}^2}$,
            \begin{equation}\label{eq_tau_ms}
                \tau_{\mathrm{ms}} = \log_2 \left( 1+ \frac{\eta P_\mathrm{Tm}}{\sigma^2 W_\mathrm{m} (\theta_\mathrm{m}+1) {D_{\mathrm{ms}}}^{\alpha_\mathrm{m}}} \right),
            \end{equation}
    and $D_{\mathrm{ms}}$ denotes the distance between the MBS and SC.
    \hfill \rule{4pt}{8pt}

    $\frac{R_\mathrm{th}}{w_{\mathrm{ss}}} \rightarrow 0$ and $\frac{R_\mathrm{th}}{w_{\mathrm{ms}}}\rightarrow 0$ hold when the data rate requirement is relatively low compared with the spectrum resource, which is reasonable since BSs usually serve many users at the same time in practical systems.
    Eqs.~(\ref{eq_outage_SC_simple}) and (\ref{eq_outage_ms_simple}) suggest that the average data rate of the non-cell-edge users should be no smaller than $R_\mathrm{th}$.
    As the proof of Lemma~2 is similar to that of Lemma~1, the details are omitted.
    We evaluate the accuracy of Lemmas~1 and 2 in Section~\ref{sec_numerical_results}.
    In addition, the outage probability constraint of the MMUs can also be simplified in the same way as Lemma~1, by approximating the distribution of MMUs as a PPP with density $\rho'_\mathrm{m} = \rho_\mathrm{m} ({D_\mathrm{m}}^2-{D_\mathrm{s}}^2)/{D_\mathrm{m}}^2$.
    The constraint $G_\mathrm{mm} =\mathds{P}\{r_\mathrm{mm}<R_\mathrm{th}\} \leq \eta$ is equivalent to
            \begin{equation}\label{eq_outage_MBS_simple}
                \bar{w}_\mathrm{mm} \tau_\mathrm{mm} \geq R_\mathrm{th},
            \end{equation}
    where $\bar{w}_\mathrm{mm}=\frac{w_\mathrm{mm}}{ 1+ \pi {D_\mathrm{m}}^2 \rho'_\mathrm{m} }$ and $\tau_\mathrm{mm}$ are given by
            \begin{equation}\label{eq_tau_MBS}
                \tau_\mathrm{mm} = \log_2 \left( 1+ \frac{P_\mathrm{Tm}}{\sigma^2 W_\mathrm{m} (\theta_\mathrm{m}+1)} \frac{\alpha_\mathrm{m}+2}{2 W_\mathrm{m}} \frac{\eta}{{D_\mathrm{m}}^{\alpha_\mathrm{m}}} \right).
            \end{equation}

    \subsection{On-Grid Power Consumption}

    Based on the simplified outage probability, we derive the on-grid power consumptions when the SC is active or turned off, respectively.
    According to Eq.~(\ref{eq_mu}), the bandwidth utilized at the SC $w_\mathrm{ss}$ is constrained by the energy consumption rate:
        \begin{equation}\label{eq_RSC_ws}
            w_\mathrm{ss} = \frac{1}{\beta_\mathrm{s} W_\mathrm{s}} \left(\mu_\mathrm{E} E - P_{0\mathrm{s}}\right).
        \end{equation}
    Based on Lemma~1, we obtain the maximum offloading ratio:
        \begin{equation}\label{eq_ratio_offloaded}
            \varphi = \frac{\frac{\tau_{\mathrm{ss}} w_\mathrm{ss} }{R_\mathrm{th}}-1}{\rho_\mathrm{s} \pi {D_\mathrm{s}}^2}.
        \end{equation}
    Besides, according to Eq.~(\ref{eq_outage_ms_simple}), the minimum bandwidth needed at the MBS to serve the MSUs is given as follows:
        \begin{equation}\label{eq_w_msa}
            w_\mathrm{msa} = \frac{R_\mathrm{th}}{\tau_\mathrm{ms}} \left( 1+ (1-\varphi) \rho_\mathrm{s} \pi {D_\mathrm{s}}^2 \right).
        \end{equation}
    However, the SC may be shut down due to empty battery, and meanwhile extra bandwidth $w_\mathrm{mso}$ is utilized at the MBS to serve the SSUs.
    Similar to Eq.~(\ref{eq_w_msa}), we obtain $w_\mathrm{mso}$:
        \begin{equation}\label{eq_w_mso}
            w_\mathrm{mso} = \frac{R_\mathrm{th}}{\tau_\mathrm{ms}} \left( 1+ \varphi \rho_\mathrm{s} \pi {D_\mathrm{s}}^2 \right).
        \end{equation}
    Therefore, the average bandwidth utilized at the MBS is given by $w_\mathrm{mm} + w_\mathrm{msa} + p_\mathrm{off} w_\mathrm{mso}$, and the average on-grid power consumption is
        \begin{equation}\label{eq_P_sum_o}
            P_\mathrm{MBS}^{\mathrm{a}} = P_{0\mathrm{m}} + \beta_\mathrm{m} \frac{P_\mathrm{Tm}}{W_\mathrm{m}} \left( w_\mathrm{mm} + w_\mathrm{msa} + p_\mathrm{off} w_\mathrm{mso} \right) + P_\mathrm{ho},
        \end{equation}
    where $p_\mathrm{off}$ and $P_\mathrm{ho}$ are given by Eq.~(\ref{eq_MD1_steady}) and Eq.~(\ref{eq_RSC_switching}) when $\lambda_\mathrm{E}< \mu_\mathrm{E}$, i.e., the energy queue is stable.
    Otherwise, the renewable energy is sufficient, $p_\mathrm{off}=0$ and $P_\mathrm{ho}=0$.
    By combining Eqs.~(\ref{eq_RSC_ws})-(\ref{eq_P_sum_o}), the average on-grid power consumption can be derived when the SC is applied for offloading, with respect to the energy consumption rate of the SC $\mu_\mathrm{E}$.

    When the SC is turned off completely, $\varphi=0$, and the bandwidth utilized at the MBS to serve the MSUs is equivalent to $w_\mathrm{msa} + w_\mathrm{mso}$.
    Thus, the on-grid power consumption is 
        \begin{equation}\label{eq_P_sum_a}
            P_\mathrm{MBS}^{\mathrm{o}} = P_{0\mathrm{m}} + \beta_\mathrm{m} \frac{P_\mathrm{Tm}}{W_\mathrm{m}} \left( w_\mathrm{mm} + w_\mathrm{msa} + w_\mathrm{mso} \right).
        \end{equation}
    Therefore, the on-grid power saving gain through offloading traffic from the MBS to the SC can be obtained:
        \begin{equation}\label{eq_delta_P}
            \begin{split}
                \Delta P_\mathrm{MBS} & = P_\mathrm{MBS}^{\mathrm{o}} - P_\mathrm{MBS}^{\mathrm{a}} \\
                & = \frac{P_\mathrm{Tm}}{W_\mathrm{m}} \beta_\mathrm{m} w_\mathrm{mso} (1-p_\mathrm{off}) - P_\mathrm{ho}\\
                & \triangleq \Delta P_\mathrm{MBS}^\mathrm{RF} - P_\mathrm{ho},
            \end{split}
        \end{equation}
     where the first term $\Delta P_\mathrm{MBS}^\mathrm{RF}$ represents the radio frequency power saved at the MBS.
     Based on Eqs.~(\ref{eq_RSC_ws}-\ref{eq_w_mso}), we have
        \begin{equation}\label{eq_gain_MBS_g} \footnotesize
            \Delta P_\mathrm{MBS}^\mathrm{RF} = \left\{ \begin{array}{ll} \zeta_\mathrm{EE} \mu_\mathrm{E} E - \left( \zeta_\mathrm{EE} P_\mathrm{0s} + \frac{\beta_\mathrm{m} P_\mathrm{Tm} R_\mathrm{th}}{W_\mathrm{m} \tau_\mathrm{ms}} \right), &  \mu_\mathrm{E}\leq \lambda_\mathrm{E} \\
            \zeta_\mathrm{EE} \lambda_\mathrm{E} E - \frac{\lambda_\mathrm{E}}{\mu_\mathrm{E}} \left( \zeta_\mathrm{EE} P_\mathrm{0s} + \frac{\beta_\mathrm{m} P_\mathrm{Tm} R_\mathrm{th}}{W_\mathrm{m} \tau_\mathrm{ms}} \right), & \mu_\mathrm{E} > \lambda_\mathrm{E}
        \end{array} \right. ,
        \end{equation}
    which $\zeta_\mathrm{EE} = \frac{W_\mathrm{s}\tau_\mathrm{ss} \beta_\mathrm{m} P_\mathrm{Tm}} {W_\mathrm{m} \tau_\mathrm{ms} \beta_\mathrm{s} P_\mathrm{Ts}}$.
    In fact, Eq.~(\ref{eq_gain_MBS_g}) reflects the conversion rate of harvested energy into on-grid power, i.e., the linear relationship between $\Delta P_\mathrm{MBS}^\mathrm{RF}$ and $\lambda_\mathrm{E}$.
    Substituting $\Delta P_\mathrm{MBS}^\mathrm{RF}$ and $P_\mathrm{ho}$ in Eq.~(\ref{eq_delta_P}) with Eqs.~(\ref{eq_gain_MBS_g}) and (\ref{eq_RSC_switching}), the power saving gain $\Delta P_\mathrm{MBS}$ can be obtained.
    We summarize the relationship between $\Delta P_\mathrm{MBS}$ and $\mu_\mathrm{E}$ in Proposition~1.

    \emph{\textbf{Proposition~1.}} Let $\kappa = \zeta_\mathrm{EE} P_\mathrm{0s} + \frac{\beta_\mathrm{m} P_\mathrm{Tm} R_\mathrm{th}}{W_\mathrm{m} \tau_\mathrm{ms}}$, then $\Delta P_\mathrm{MBS}$ has following properties:
        \begin{enumerate}
            \item $\Delta P_\mathrm{MBS}$ increases linearly with $\mu_\mathrm{E}$ for $\mu_\mathrm{E}\leq \lambda_\mathrm{E}$;
            \item $\Delta P_\mathrm{MBS}$ increases with $\mu_\mathrm{E}$ if $\kappa \geq 3 \lambda_\mathrm{E} C_\mathrm{ho}$ ;
            \item $\Delta P_\mathrm{MBS}$ decreases with $\mu_\mathrm{E}$ for $\mu_\mathrm{E} > \lambda_\mathrm{E}$ if $\kappa \leq (1-\frac{1}{e}) \lambda_\mathrm{E} C_\mathrm{ho}$ ;
            \item If $(1-\frac{1}{e}) \lambda_\mathrm{E} C_\mathrm{ho} < \kappa < 3 \lambda_\mathrm{E} C_\mathrm{ho}$, $\Delta P_\mathrm{MBS}$ is a concave function of $\frac{\lambda_\mathrm{E}}{\mu_\mathrm{E}}$ for $\mu_\mathrm{E} > \lambda_\mathrm{E}$, and the optimal condition is
                    \begin{equation}\label{eq_opt_mu_RSC}\footnotesize
                        \lambda_\mathrm{E}C_\mathrm{ho} \frac{e^{-\frac{\lambda_\mathrm{E}}{\mu_\mathrm{E}}}}{\left(\frac{\lambda_\mathrm{E}}{\mu_\mathrm{E}}\right)^2} \left( -e^{-\frac{\lambda_\mathrm{E}}{\mu_\mathrm{E}}} +1+\frac{\lambda_\mathrm{E}}{\mu_\mathrm{E}}-\left(\frac{\lambda_\mathrm{E}}{\mu_\mathrm{E}}\right)^2 \right) =  \kappa.
                    \end{equation}
        \end{enumerate}
    \emph{Proof}: Please refer to Appendix~\ref{appendix_RSC_gain}.
    \hfill \rule{4pt}{8pt}

    As $0\leq w_\mathrm{ss}\leq W_\mathrm{s}$ and $\varphi \leq 1$, $\mu_\mathrm{E}$ should satisfy
        \begin{equation}\label{eq_mu_condition}\footnotesize
            \frac{P_{0\mathrm{s}}}{E} \leq \mu_\mathrm{E} \leq \frac{1}{E} \left\{ P_{0\mathrm{s}}+ \min \left\{1, \frac{R_\mathrm{th}}{\tau_\mathrm{ss} W_\mathrm{s}}\left(\rho_\mathrm{s}\pi D_\mathrm{s}^2 +1 \right) \right\} \beta_\mathrm{s} P_\mathrm{Ts} \right\}.
        \end{equation}
    Furthermore, the optimal energy consumption rate $\tilde{\mu}_\mathrm{E}$ can be obtained by combining Proposition~1 and Eq.~(\ref{eq_mu_condition}).
    For example, the SC should be active and work at the maximal energy consumption rate when $C_\mathrm{ho}=0$ (i.e., no handover cost), according to the second property of Proposition~1.
    In addition, the SC should be activated only when the power saving gain is positive; otherwise, the activation will increase the on-grid power consumption.

    At this point, the energy-optimal traffic shaping (EOTS) scheme has been obtained, whereby the on-off state of the SC and the amount of traffic offloaded should be dynamically adjusted with the energy arrival rate, based on Proposition~1 and Eq.~(\ref{eq_mu_condition}).

\begin{table}[!t]
        \caption{Simulation Parameters \cite{EARTH}}
        \label{tab_parameter}
        \centering
        \begin{tabular}{cccc}
        \hline
        \hline
        Parameter & Value & Parameter & Value \\
        \hline
        $D_\mathrm{m}$ & 1000m & $D_\mathrm{s}$ & 300m \\
        $P_{0\mathrm{m}}$ & 130W & $P_{0\mathrm{s}}$ & 56W \\
        $P_{\mathrm{Tm}}$ & 20W & $P_{\mathrm{Ts}}$ & 6.3W \\
        $\beta_\mathrm{m}$ & 4.7 & $\beta_\mathrm{s}$ & 2.6 \\
        $\alpha_\mathrm{m}$ & 3.5 & $\alpha_\mathrm{s}$ & 4 \\
        $W_\mathrm{m}$ & 10MHz & $W_\mathrm{s}$ & 10MHz\\
        $R_\mathrm{th}$ & 100kbps & $\eta$ & 0.05\\
        $\theta_\mathrm{m}$ & 1000 & $\theta_\mathrm{s}$ & 2000\\
        $\sigma$ & -105dBm/MHz &  $D_\mathrm{ms}$ & 600m\\
        \hline
        \hline
        \end{tabular}
\end{table}

\begin{figure}
    \centering
  \includegraphics[width=2.5in]{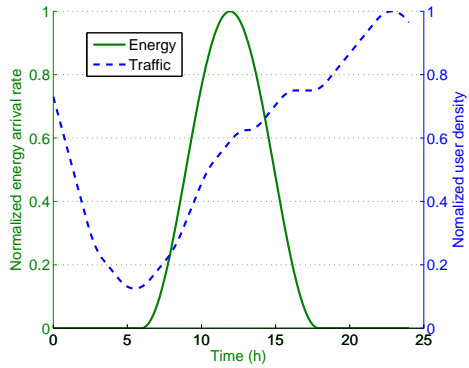}\\
  \caption{Daily traffic and energy profiles.}\label{fig_traffic_energy}
\end{figure}

%%%%%%%%%%%%%%%%%%%%%%%%%%%%%%%%%%%%%%%%%%%%%%%%%%%%%%%%%%%%%%%%%%%%%%%%%%%%%%%%%%%%%%%%%%%%%%%%%%%
\section{Simulation Results}
    \label{sec_numerical_results}

In this section, the accuracy of the derived outage probability in Lemmas~1 and 2 is validated, and the power saving performance of EOTS is also evaluated.
Simulation parameters are listed in Table~\ref{tab_parameter}.
Solar powered SC is considered.
Fig.~\ref{fig_traffic_energy} shows the daily traffic \cite{EARTH} and solar energy profiles \cite{JGong_TC}. %\cite{solar_sine}

 \begin{figure}[!t]
        \centering
        \subfloat[MMU and SSU] {\includegraphics[width=2.5in]{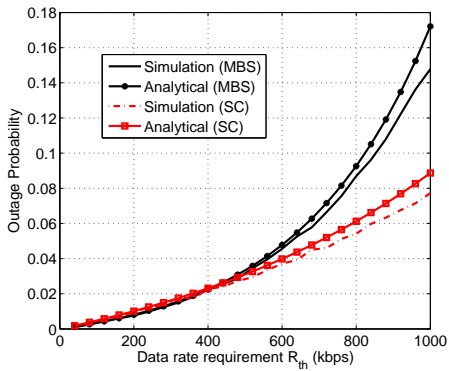}
        \label{fig_evaluation_MU_SU}}
        \hfil\\
        \subfloat[MSU]{\includegraphics[width=2.5in]{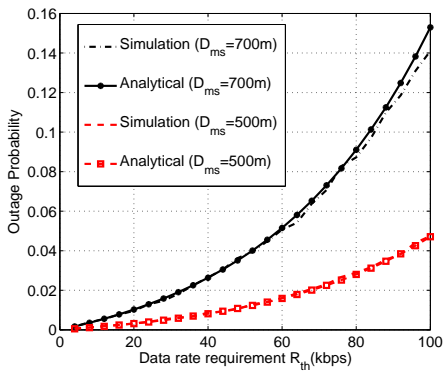}
        \label{fig_evaluation_MSU}}
        \caption{Outage probability.}
        \label{fig_evaluation}
    \end{figure}

\subsection{Outage Probability Evaluation}

We evaluate the analytical results in Lemmas~1 and 2 via Monte Carlo simulations, with user density set as $\rho_\mathrm{m}$=20/km$^2$ and $\rho_\mathrm{s}$=60/km$^2$.
The number of users and their locations are generated randomly, and the results of 5000 simulation samples are averaged.
The analytical results obtained by Eqs.~(\ref{eq_outage_SC_simple}) and (\ref{eq_outage_MBS_simple}) are compared with the simulation ones in Fig.~\ref{fig_evaluation}(a), which are shown to increase with the data rate requirement.
Fig.~\ref{fig_evaluation}(b) validates Lemma~2 with different distances between the MBS and the SC, when the bandwidth allocated to MSUs is set as $w_\mathrm{ms}=1$ MHz.
It can be seen that the analytical and simulation results match well, with acceptable deviations.

\begin{figure}[!t]
        \centering
        \includegraphics[width=2.5in]{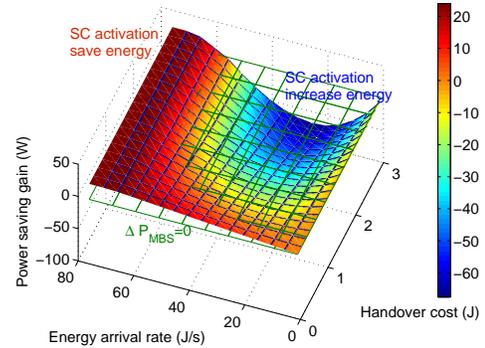}
        \caption{Maximal power saving gain of traffic offloading.}
        \label{fig_gain_single_SC}
    \end{figure}

\begin{figure}[!t]
        \centering
        \includegraphics[width=2.5in]{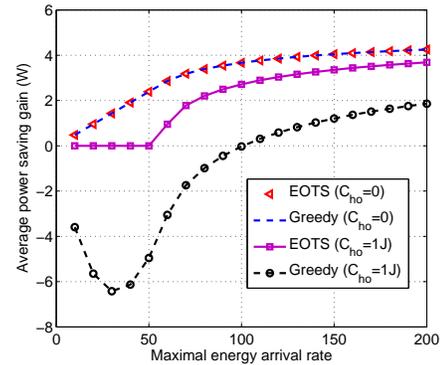}
        \caption{Average power saving gain of EOTS.}
        \label{fig_gain_single_SC_daily}
    \end{figure}

\subsection{Power Saving Gain through Traffic Shaping}

Fig.~\ref{fig_gain_single_SC} shows the power saving gain achieved by activating the SC for traffic offloading, with the optimal energy consumption rate.
Firstly, the power saving gain decreases with the handover cost $C_\mathrm{ho}$, which is straightforward.
Secondly, the power saving gain firstly decreases and then increases with the energy arrival rate.
The reason is that the on-grid power consumption is dominated by the handover cost when the energy arrival rate $\lambda_\mathrm{E}$ is small, due to the insufficient green energy supply.
Besides, the handover frequency increases with the energy arrival rate, given that the energy queue is empty during most of the time with low energy arrival rate.
Accordingly, the total power saving gain decreases when $\lambda_\mathrm{E}$ is small.
Then, as $\lambda_\mathrm{E}$ further increases, the harvested energy balances out the handover cost, which improves the power saving gain.

Fig.~\ref{fig_gain_single_SC_daily} shows the power saving performance of EOTS, which is compared with the greedy scheme for the traffic and energy arrival profiles given by Fig.~\ref{fig_traffic_energy}.
For the greedy scheme, the SC is always active and work at the maximal energy consumption rate.
The average power saving gain with respect to the maximum energy arrival rate is shown, where the maximal user density is set as $\rho_\mathrm{s}^\mathrm{max} = 100/$km$^2$.
When the handover cost is ignored ($C_\mathrm{ho}=0$), the two schemes have the same performance.
However, EOTS is more advantageous when the handover cost increases.
Notice that, the power saving gain of the greedy scheme may even become negative, i.e., deploying a off-grid SC may increase the total power consumption without effective traffic management.
Whereas, EOTS always reduces the on-grid power consumption.

%%%%%%%%%%%%%%%%%%%%%%%%%%%%%%%%%%%%%%%%%%%%%%%%%%%%%%%%%%%%%%%%%%%%%%%%%%%%%%%%%%%%%%%%%%%%%%%%%%%
\section{Conclusions and Future Work}
    \label{sec_conclusions}

    In this paper, we derived the closed-form expression of the on-grid power saving gain achieved by traffic offloading, whereby users are dynamically offloaded from the conventional on-grid MBS to the off-grid SC powered solely by EH.
    The derived power saving gain indicates the conversion rate of harvested energy into on-grid power, with spatial traffic shaping.
    Besides, an energy-optimal traffic shaping scheme, EOTS, is proposed, which offers a guideline for practical energy-aware network operations.
    In the future, the co-existence of multiple SCs with diverse power sources will be studied.

%%%%%%%%%%%%%%%%%%%%%%%%%%%%%%%%%%%%%%%%%%%%%%%%%%%%%%%%%%%%%%%%%%%%%%%%%%%%%%%%%%%%%%%%%%%%%%%%%%%

\appendices{}

\section{Proof of Lemma~1}
    \label{appendix_SC}

        Due to channel fading or bandwidth limitation, the achievable rate of a user may be smaller than its required threshold $R_\mathrm{th}$, and correspondingly, this rate outage probability is defined as $G_\mathrm{ss}=\mathds{P}\{r_\mathrm{ss}<R_\mathrm{th}\}$.
        Next, we derive the closed-form expression of $G_\mathrm{ss}$, based on the Eqs.~(\ref{eq_SINR_SC}) and (\ref{eq_r_s}).
        As the distribution of SSUs follows PPP with density $\varphi \rho_{\mathrm{s}}$, the probability distribution function of the distance between SSUs and the SC $d_\mathrm{s}$ is $f_{d_\mathrm{s}}(d) = \frac{2d}{D_{\mathrm{s}}^2}$.
        Based on Eq.~(\ref{eq_SINR_SC}), we have

        \begin{subequations} \footnotesize
            \label{eq_appendix_MBS_outage_Nm}
            \begin{align}
                % & \mathds{P} \left\{ \log_2 (1 + \gamma_{\mathrm{ss}}) \frac{w_\mathrm{ss}}{K_{\mathrm{ss}}+1} < R_\mathrm{th} \right\}\\
                & \mathds{P} \left\{\gamma_{\mathrm{ss}} \geq 2^{(K_{\mathrm{ss}}+1)\frac{R_\mathrm{th}}{ w_{\mathrm{ss}}}} - 1 \right\} \nonumber \\
                &= \int_0^{D_{\mathrm{s}}} \mathds{P} \left\{ h_{\mathrm{s}} \geq \frac{(\theta_\mathrm{s}+1)\sigma^2 {W_\mathrm{s}}}{P_{\mathrm{Ts}} d^{-\alpha_\mathrm{s}}} \left(2^{(K_{\mathrm{ss}}+1) \frac{R_\mathrm{th}}{w_{\mathrm{ss}}}} -1 \right)\right\} \frac{2d}{D_{\mathrm{s}}^2} \mathrm{d} d  \nonumber \\
                &= \int_0^{D_{\mathrm{s}}} \exp\left( -\frac{(\theta_\mathrm{s}+1)\sigma^2 {W_\mathrm{s}}}{P_{\mathrm{Ts}} d^{-\alpha_\mathrm{s}} } \left(2^{(K_{\mathrm{ss}}+1) \frac{R_\mathrm{th}}{w_{\mathrm{ss}}}} -1 \right)\right) \frac{2d}{D_{\mathrm{s}}^2} \mathrm{d} d
                \label{eq_appendix_MBS_rayleigh} \\
                &= \int_0^{D_{\mathrm{s}}} \left( 1-\frac{(\theta_\mathrm{s}+1)\sigma^2 {W_\mathrm{s}} }{P_{\mathrm{Ts}} d^{-\alpha_\mathrm{s}}} \left(2^{(K_{\mathrm{ss}}+1) \frac{R_\mathrm{th}}{w_{\mathrm{ss}}}} -1 \right)\right) \frac{2d}{D_{\mathrm{s}}^2} \mathrm{d} d  \label{eq_appendix_MBS_PDF_rate} \\
                &= 1 - \frac{2D_{\mathrm{s}}^{\alpha_\mathrm{s}}}{\alpha_\mathrm{s}+2} \frac{(\theta_\mathrm{s}+1)\sigma^2 {W_\mathrm{s}} } {P_{\mathrm{Ts}}} \left( 2^{(K_{\mathrm{ss}}+1) \frac{R_\mathrm{th}}{w_{\mathrm{ss}}}}-1 \right), \nonumber
            \end{align}
        \end{subequations}
    where Eq.~(\ref{eq_appendix_MBS_rayleigh}) is due to the property of Rayleigh fading, and (\ref{eq_appendix_MBS_PDF_rate}) comes from $\frac{\sigma^2 {W_\mathrm{s}}}{P_{\mathrm{Ts}}}\rightarrow 0$ (i.e., high signal to noise ratio).
    In addition, $K_{\mathrm{ss}}$ should follow the Poisson distribution of parameter $\pi \varphi \rho_{\mathrm{s}} {D_{\mathrm{s}}}^2$, according to the Slivnyak-Mecke theorem \cite{Stochastic_Geometry}.
    Therefore, the outage probability of SSU $G_\mathrm{ss}$ satisfies
        \begin{equation} \footnotesize
            \begin{split}
                & G_{\mathrm{ss}} \!= \!1\!-\!\sum_{K=0}^\infty \mathds{P} \left( \gamma_{\mathrm{ss}} \geq 2^{(K+1) \frac{R_\mathrm{th}}{w_{\mathrm{ss}}}} -1 \right) P_{K_{\mathrm{ss}}}(K) \\
                &=\! 1\!-\!\sum_{K=0}^\infty \mathds{P} \left( \gamma_{\mathrm{ss}} \geq 2^{(K+1) \frac{R_\mathrm{th}}{w_{\mathrm{ss}}}} -1 \right) \frac{(\pi D_{\mathrm{s}}^2 \rho_{\mathrm{s}})^K}{K!}e^{- \pi D_{\mathrm{s}}^2 \rho_{\mathrm{s}}} \\
                &=\frac{2 D_{\mathrm{s}}^{\alpha_\mathrm{s}} (\theta_\mathrm{s}\!+\!1)\sigma^2 {W_\mathrm{s}} }{P_{\mathrm{Ts}} \left(\alpha_\mathrm{s}\!+\!2\right) } \! \left( 2^{\frac{R_\mathrm{th}}{w_{\mathrm{ss}}}}\exp\left(\pi D_{\mathrm{s}}^2 \rho_{\mathrm{s}} \left( 2^{\frac{R_\mathrm{th}}{w_{\mathrm{ss}}}} \!-\! 1 \right)\right)\!-\!1\right).
            \end{split}
        \end{equation}
    Furthermore, $\frac{R_\mathrm{th}}{w_{\mathrm{ss}}} \rightarrow 0$ (i.e., sufficient bandwidth), $2^{\frac{R_\mathrm{th}}{w_{\mathrm{ss}}}}-1=\ln2 \cdot \frac{R_\mathrm{th}}{w_{\mathrm{ss}}}$. %and
            %\begin{equation}\label{eq_outage_SC_closed_2}\footnotesize
             %   G_{\mathrm{ss}} \!=\! \frac{2 {D_{\mathrm{s}}}^{\alpha_\mathrm{s}} (\theta_\mathrm{s}\!+\!1)\sigma^2}{P_{\mathrm{Ts}} \left(\alpha_\mathrm{s}\!+\!2\right) } \! \left(2^{\frac{R_\mathrm{th}}{w_{\mathrm{ss}}}\left(1 + \pi {D_{\mathrm{s}}}^2 \varphi_n \rho_{\mathrm{s}} \frac{R_\mathrm{th}}{w_{\mathrm{ss}}} \right)}\!-\!1\right),
            %\end{equation}
    %which is equivalent to Eq.~(\ref{eq_outage_SC_simple}).
    Hence, Lemma~1 is proved.

\section{Proof of Proposition~1}
    \label{appendix_RSC_gain}

    Denote $\varrho = \frac{\lambda_\mathrm{E}}{\mu_\mathrm{E}}$ where $\varrho\in(0,1)$ for $\mu_\mathrm{E}>\lambda_\mathrm{E}$, and we take the derivation of $\Delta P_\mathrm{MBS}$ with respect to $\varrho$:
        \begin{equation}\label{appendix_diff_delta_g}\footnotesize
            \begin{split}
                \frac{\mathrm{d} \Delta P_\mathrm{MBS}}{\mathrm{d} \varrho} % & = - \kappa - \left[ (1-\frac{1}{\varrho^2})(1-e^{-\varrho}) +(\frac{1}{\varrho}-1)e^{-\varrho} \right] \lambda_\mathrm{E} C_\mathrm{ho}\\
                & = - \kappa - \lambda_\mathrm{E} C_\mathrm{ho} \frac{e^{-\varrho}}{\varrho^2} \left( -e^\mathrm{\varrho} +1+\varrho-\varrho^2 \right), \\
                & \triangleq - \kappa + \lambda_\mathrm{E} C_\mathrm{ho} f(\varrho).
            \end{split}
        \end{equation}
    Notice that
        \begin{subequations}\label{appendix_fx}\footnotesize
            \begin{align}
                \frac{\mathrm{d}f(\varrho)}{\mathrm{d} \varrho} % & = \frac{2 e^{-\varrho}}{\varrho^3} \left( -e^{\varrho} +1+\varrho+\frac{\varrho^2}{2} - \frac{\varrho^3}{2}  \right)\\
                & = \frac{2 e^{-\varrho}}{\varrho^3} \left( - \frac{\varrho^3}{2} - \sum\limits_{i=3}\limits^{\infty} \frac{\varrho^i}{i!}  \right) <0,
            \end{align}
        \end{subequations}
    we have $ 1-e^{-1} < f(\varrho) < 3/2$ for $0\leq \varrho \leq 1$.
    In addition, $\Delta P_\mathrm{MBS}$ is concave with respect to $\varrho$, as $\frac{\mathrm{d}^2 \Delta P_\mathrm{MBS}}{\mathrm{d} \varrho^2} = f'(\varrho)<0$.
    Specifically, $\frac{\mathrm{d} \Delta P_\mathrm{MBS}}{\mathrm{d} \varrho} < 0$ for $\kappa \geq 3\lambda_\mathrm{E}C_\mathrm{ho}$, and $\frac{\mathrm{d} \Delta P_\mathrm{MBS}}{\mathrm{d} \varrho} > 0$ for $\kappa \leq (1-\frac{1}{e}) \lambda_\mathrm{E}C_\mathrm{ho}$.
    Otherwise, there exists $\tilde{\varrho}$ satisfying $ \frac{\mathrm{d} \Delta P_\mathrm{MBS}}{\mathrm{d} \varrho}|_{\tilde{\varrho}} = 0$, and the corresponding energy consumption rate $\mu_\mathrm{E} = \frac{\lambda_\mathrm{E}}{\tilde{\varrho}}$ can maximize $\Delta P_\mathrm{MBS}$. Hence, Proposition~1 is proved.

%figures
%    \clearpage
%    \input{figure_double.tex}

\end{document}